\documentclass{aa}
\usepackage[T1]{fontenc}
\usepackage{color}
\usepackage{graphicx}
\usepackage[usenames,dvipsnames]{xcolor}
\usepackage{natbib}
\usepackage[varg]{txfonts}
\usepackage{longtable}
\usepackage[hyperfootnotes=true, hidelinks, colorlinks, citecolor=blue, linkcolor=blue, linktocpage, bookmarks=true]{hyperref}
\usepackage{footnote}
\usepackage{float}
\usepackage{multirow}
\usepackage{stackengine}
\usepackage{amsmath}
\usepackage{lscape}

\begin{document}

\title{Improving the spin-down limits of the continuous gravitational waves emitted from rotating triaxial pulsars}
\titlerunning{Improving the spin-down limits}

\author{D. Pathak\inst{1}
    \and D. Chatterjee\inst{1}
    }

\institute{Inter-University Centre for Astronomy and Astrophysics, Pune University Campus, Pune - 411007, India \\ \email{dhruv.pathak@iucaa.in}
}

\date{Received Date Month Year / Accepted Date Month Year}

\abstract{The spin-down limit of the continuous gravitational wave strain from pulsars assumed to be triaxial stars rotating about a principal moment of inertia axis depends upon the value of the intrinsic spin frequency derivative of the pulsar, among other parameters. In order to get more accurate intrinsic spin frequency derivative values, dynamical effects contributing to the measured spin frequency derivative values must be estimated via more realistic approaches. In this work, we calculated improved values for the spin-down limit of the continuous gravitational wave strain (assuming that pulsars are triaxial stars rotating about a principal moment of inertia axis) for a set of 237 pulsars for which a targeted search for continuous gravitational waves was recently carried out by the LIGO-Virgo-KAGRA (LVK) Collaboration. We used `GalDynPsr', a Python-based public package, to calculate more realistic values for the intrinsic spin frequency derivatives and, consequently, we get more realistic values of the spin-down limit. The realistic values that we obtain for the intrinsic spin frequency derivatives can also provide a valuable contribution to improving the sensitivity of searches for continuous gravitational waves from known pulsars.}

\maketitle

\section{Introduction}

The measured values of pulsar parameters, such as spin frequency and its derivatives ($f_{\rm s}$, $\dot{f}_{\rm s}$), are affected by the pulsar's velocity and acceleration. These measured values hence contain contributions from additional terms that are functions of a pulsar's velocity and acceleration (and even their derivatives) and are called the dynamical effect terms. A source of these dynamical effects that is common to all the pulsars in our Galactic field is the gravitational potential of the Milky Way \citep{pb18}.

While calculating parameters like the braking index ($n$) and the spin-down limit of the gravitational wave strain ($h_{0}^{\rm sd}$), it is essential to use the intrinsic values of frequency derivatives, that is, ones that are devoid of any contamination from dynamical effects, in order to get more accurate results for these parameters.

The spin-down limit of the continuous gravitational wave strain from a pulsar assumed to be a triaxial star rotating about a principle moment of inertia axis is expressed as
\begin{align}
h_{0}^{\rm sd}=\frac{1}{d}\left(\frac{5G I_{zz} |\dot{f}_{\rm s,int}|}{2 c^3 f_{\rm s}}\right)^{1/2},    
\label{eq:h0sd}
\end{align}
where $\dot{f}_{\rm s,int}$ is the intrinsic value of the spin frequency derivative, $c$ is the speed of light in vacuum, $G$ is the gravitational constant, $d$ is the distance to the pulsar, and $I_{zz}$ is the z component of the moment of inertia.

Recently, \citet{aa21} gave $h_{0}^{\rm sd}$ values for 237 real pulsars. However, they only corrected for dynamical effects when calculating $\dot{f}_{\rm s,int}$ values for 144 pulsars out of the 237, and even here used a traditional approach \citep{dt91} that does not work for pulsars further away from the Sun. In Sect. 2, we discuss the conventional approaches and their shortcomings, and suggest a more realistic approach to estimating the dynamical corrections to the spin period (or frequency) derivative of pulsars in order to estimate their intrinsic value, which is used in the calculation of $h_{0}^{\rm sd}$ values. In Sect. 3, we present our improved $h_{0}^{\rm sd}$ values based on $\dot{f}_{\rm s,int}$ values calculated using a Python-based package, `GalDynPsr' \citep{pb18}, for the real pulsars in \citet{aa21}.

\section{Estimating $\dot{f}_{\rm s,int}$}

One can estimate $\dot{f}_{\rm s,int}$ from $\dot{P}_{\rm s,int}$ by following the relation
\begin{align}
\dot{f}_{\rm s,int} = -\frac{\dot{P}_{\rm s,int}}{P_{\rm s}^2} ,
    \label{eq:fpdot}
\end{align}
where $\dot{P}_{\rm s,int}$ is the intrinsic value of the spin period derivative and $P_{\rm s}$ is the spin period.

$\dot{P}_{\rm s,int}$ can be estimated by subtracting from $\left( \frac{\dot{P}_{\rm s} }{ {P}_{\rm s}} \right)_{\rm obs}$ the dynamical contributions to the observed value of the period derivative, $\left( \frac{\dot{P}_{\rm s} }{ {P}_{\rm s}} \right)_{\rm excess}$ (also called the excess term; see \citet{pb18}), as shown in the following equation:
\begin{align}
\dot{P}_{\rm s,int} = P \left(\left( \frac{\dot{P}_{\rm s} }{ {P}_{\rm s}} \right)_{\rm obs} - \left( \frac{\dot{P}_{\rm s} }{ {P}_{\rm s}} \right)_{\rm excess}\right)~,
    \label{eq:pdot1}
\end{align}
where we define the excess term, $\left( \frac{\dot{P}_{\rm s} }{ {P}_{\rm s}} \right)_{\rm excess}$, as
\begin{align}
\left( \frac{\dot{P}_{\rm s} }{ {P}_{\rm s}} \right)_{\rm excess} = \left( \frac{\dot{P}_{\rm s} }{ {P}_{\rm s}} \right)_{\rm excess, Galpl}  + \left( \frac{\dot{P}_{\rm s} }{ {P}_{\rm s}} \right)_{\rm excess, Galz}  + \left( \frac{\dot{P}_{\rm s} }{ {P}_{\rm s}} \right)_{\rm excess, Shk} ~.
\label{eq:doppler3}
\end{align}
Here, $\left( \frac{\dot{P}_{\rm s} }{ {P}_{\rm s}} \right)_{\rm excess, Galpl}$ denotes the component of the excess term that contains the contribution from the component of relative acceleration parallel to the Galactic plane, $\left( \frac{\dot{P}_{\rm s} }{ {P}_{\rm s}} \right)_{\rm excess, Galz}$ denotes the component of the excess term that contains the contribution from the component of relative acceleration perpendicular to the Galactic plane, and $\left( \frac{\dot{P}_{\rm s} }{ {P}_{\rm s}} \right)_{\rm excess, Shk}$ denotes the component of the excess term that contains the contribution from the proper motion of the pulsar (also called the Shklovskii effect term \citep{shk70}). These components of the excess term are expressed as \citep{pb18}

\begin{align}
\left( \frac{\dot{P}_{\rm s} }{ {P}_{\rm s}} \right)_{\rm excess, Galpl}  = -\frac{1}{c}  \left( {a}_{\rm p, Galpl} \, \cos \lambda + {a}_{\rm s, Galpl} \, \cos l \right)\, \cos b ~, 
\label{eq:excessGalplog}
\end{align}

\begin{align}
 \left( \frac{\dot{P}_{\rm s} }{ {P}_{\rm s}} \right)_{\rm excess, Galz} = - \frac{ 1 }{c} \, | {a}_{\rm p, Galz} | \, \sin |b| ~,
\label{eq:excessGalz1}
\end{align} 

and 

\begin{align}
 \left( \frac{\dot{P}_{\rm s} }{ {P}_{\rm s}} \right)_{\rm excess, Shk} =  \frac{1}{c} \frac{V_T^2}{d} =  2.42925 \times 10^{-21}  \, d_{\rm kpc} \, \mu_{T, {\rm mas \, yr^{-1}}}^2 ~ ~{\rm s^{-1}} \, .
\label{eq:excessSh2}
\end{align} 

In the above equations, $l$ is the Galactic longitude, $b$ is the Galactic latitude, $d$ is the distance between the pulsar and the Solar System barycentre, ${a}_{\rm p, Galpl}$ is the component of the pulsar's acceleration parallel to the Galactic plane, ${a}_{\rm s, Galpl}$ is the component of the Sun's acceleration parallel to the Galactic plane, ${a}_{\rm p, Galz}$ is the component of the pulsar's acceleration perpendicular to the Galactic plane, $d_{\rm kpc}$ is the distance in kpc, $V_T$ is the total transverse relative velocity magnitude, and $\mu_{T, {\rm mas \, yr^{-1}}}$ is the total proper motion in ${\rm mas \, yr^{-1}}$. $\cos \lambda$ is given by the relation $\cos \lambda = \frac{R_{\rm s}}{R_{\rm p}} \left( \frac{d \cos b}{R_{\rm s}} - \cos l\right)$, in which $R_{\rm s}$ is the Galactocentric cylindrical distance of the Sun and $R_{\rm p}$ is the Galactocentric cylindrical distance of the pulsar, given by $R_{\rm p}^2 = R_{\rm s}^2 + (d \cos b)^2 - 2 R_{\rm s} (d \cos b) \cos l $ \citep{pb18}.

Conventionally, when estimating $\left( \frac{\dot{P}_{\rm s} }{ {P}_{\rm s}} \right)_{\rm excess, Galpl}$, ${a}_{\rm p, Galpl}$ and ${a}_{\rm s, Galpl}$ are taken to be centripetal accelerations, with ${a}_{\rm p, Galpl}=\frac{v_{\rm p}^2}{R_{\rm p}}$ and ${a}_{\rm s, Galpl}=\frac{v_{\rm s}^2}{R_{\rm s}}$, where $v_{\rm p}$ is the Galactocentric rotational speed of the pulsar and $v_{\rm s}$ is the Galactocentric rotational speed of the Sun parallel to the Galactic plane \citep{dt91}. Furthermore, $v_{\rm p}$ is conventionally assumed to be a linear function of $R_{\rm p}$ with a negligible slope. In other words, pulsars are assumed to follow the flat rotation curve approximation \citep{dt91,rm14}. This leads to the use of an approximate expression,

\begin{align}
 \left( \frac{\dot{P}_{\rm s} }{ {P}_{\rm s}} \right)_{\rm excess, Galpl}  = -\frac{1}{c} \frac{v_{\rm s}^2}{R_{\rm s}} \left(  \cos l +  \frac{\beta}{(\sin^2 l + \beta^2)} \right)\, \cos b~,
\label{eq:excessGaldt91present}
\end{align}
where $\beta= \frac{d \cos b}{R_{\rm s}} - \cos l$. However, the flat rotation curve approximation only works for the pulsars close to the Sun, that is, those which have $R_{\rm p}$ close to $R_{\rm s}$, as the rotation curve is approximately flat for those $R_{\rm p}$ values. As we move away from the Sun, either towards the Galactic centre or towards the edges of the Galaxy, the rotation curve cannot be approximated by a flat line; hence, any estimation of $\left( \frac{\dot{P}_{\rm s} }{ {P}_{\rm s}} \right)_{\rm excess, Galpl}$ using the flat rotation curve approximation is bound to give inaccurate results for such pulsars \citep{pb18}. Additionally, in Eq. (\ref{eq:excessGaldt91present}), we can see that the dependence on the $z$ value, that is, the vertical separation of the pulsar from the Galactic plane, is not taken into account.

Similarly, when estimating $\left( \frac{\dot{P}_{\rm s} }{ {P}_{\rm s}} \right)_{\rm excess, Galz}$, the approximate expressions for ${a}_{\rm p, Galz}$, given by \citet{nt95} as well as \citet{lwz09}, have conventionally been used. In these expressions, ${a}_{\rm p, Galz}$ only depends upon the $z$ value and its magnitude tends to increase with the increase in vertical separation \citep{pb18}. This, we know, is counterintuitive, as moving away from the source of gravitational potential should eventually lead to a decrease in acceleration.

We know that acceleration is essentially the gradient of the gravitational potential. In a more realistic approach, therefore, the acceleration components in Eqs. (\ref{eq:excessGalplog}) and (\ref{eq:excessGalz1}) should be expressed in terms of gradients of the gravitational potential of the Milky Way. Specifically, ${a}_{\rm p, Galpl}$ and ${a}_{\rm s, Galpl}$ should be estimated as the gradient of the Galactic potential along the Galactocentric cylindrical radial directions at the location of the pulsar and the Sun, respectively. ${a}_{\rm p, Galz}$, on the other hand, should be estimated as the gradient of the Galactic potential along the z direction at the location of the pulsar.

One such model for the Galactic potential is given in the publicly available Python package `galpy' \citep{bovy15}. It contains contributions from the Galactic disk, the central bulge, the dark matter halo, and the central supermassive black hole, and is referred to as `MWPotential2014wBH'. galpy also provides other Milky Way potential models like `MWPotential2014' (the same as `MWPotential2014wBH' except that it does not incorporate the contribution from the central supermassive black hole) and `McMillan17' (which is based on the Milky Way potential given by \citet{mc17}). However, the values of acceleration components calculated using these different Milky Way potential models are in agreement with each other for real pulsars (for details, see \citet{pb18,bo20}). In this work, we used Model-Lb of GalDynPsr, which uses the `MWPotential2014wBH' model from galpy for acceleration calculations, because, in addition to contributions from the Galactic disk, the central bulge, and the dark matter halo, we wanted to add the contribution from the central supermassive black hole for completeness. We discuss this in more detail in the following section.

\section{Improving the $h_{0}^{\rm sd}$ values of real pulsars: Correcting for dynamical effects in frequency derivatives}

Tables 3 and 4 of \citet{aa21} jointly give a list of $h_{0}^{\rm sd}$ values for 237 real pulsars. They also provide the values of $f_{\rm s}$ and $\dot{P}_{\rm s}$, which are used in the calculation of $h_{0}^{\rm sd}$ values. Out of these 237 pulsars, they corrected for the dynamical effects for only 144 pulsars. However, for these 144 pulsars, the dynamical effects corrected for include the Shklovskii effect (as given in Eq. (\ref{eq:excessSh2})) and the Galactic rotation effect (as given in Eq. (\ref{eq:excessGaldt91present})), and hence, the $ \left( \frac{\dot{P}_{\rm s} }{ {P}_{\rm s}} \right)_{\rm excess, Shk}$ and $ \left( \frac{\dot{P}_{\rm s} }{ {P}_{\rm s}} \right)_{\rm excess, Galpl}$ values were calculated but the $ \left( \frac{\dot{P}_{\rm s} }{ {P}_{\rm s}} \right)_{\rm excess, Galz}$ values were not corrected for in \citet{aa21}. Moreover, the $ \left( \frac{\dot{P}_{\rm s} }{ {P}_{\rm s}} \right)_{\rm excess, Galpl}$ values were calculated using a flat rotation curve approximation, as given in \citet{dt91}, which, as mentioned in the previous section, is not an accurate approach as it does not work for pulsars much further away from the Sun \citep{pb18}. 

We therefore decided to use a Python package, `GalDynPsr'\footnote{https://github.com/pathakdhruv/GalDynPsr} \citep{pb18}, to estimate more realistic $\dot{f}_{\rm s,int}$ (intrinsic spin frequency derivative) values for all the pulsars given in \citet{aa21}. We specifically used Model-Lb in GalDynPsr to calculate $ \left( \frac{\dot{P}_{\rm s} }{ {P}_{\rm s}} \right)_{\rm excess}$ and consequently $\dot{f}_{\rm s, int}$. This model incorporates the use of the Milky Way potential `MWPotential2014wBH', as given in galpy, to estimate the acceleration components required to calculate $ \left( \frac{\dot{P}_{\rm s} }{ {P}_{\rm s}} \right)_{\rm excess, Galpl}$ and $ \left( \frac{\dot{P}_{\rm s} }{ {P}_{\rm s}} \right)_{\rm excess, Galz}$. GalDynPsr also calculates the Shklovskii effect contribution, that is, $ \left( \frac{\dot{P}_{\rm s} }{ {P}_{\rm s}} \right)_{\rm excess, Shk}$. Consequently, after having calculated the total dynamical contribution, that is, $ \left( \frac{\dot{P}_{\rm s} }{ {P}_{\rm s}} \right)_{\rm excess, Shk}$ + $ \left( \frac{\dot{P}_{\rm s} }{ {P}_{\rm s}} \right)_{\rm excess, Galpl}$ + $ \left( \frac{\dot{P}_{\rm s} }{ {P}_{\rm s}} \right)_{\rm excess, Galz}$, GalDynPsr calculates the intrinsic value of the spin period derivative. In order to calculate $\dot{f}_{\rm s, int}$ using GalDynPsr, we required the values of the observable parameters: Galactic longitude ($l$), Galactic latitude ($b$), distance ($d$), proper motion in the right ascension direction ($\mu_\alpha$), proper motion along the declination ($\mu_\delta$), spin period ($P_{\rm s}$), and the observed value of the spin period derivative ($\dot{P}_{\rm s, obs}$) \citep{pb18}.

For the observables $l$, $b$, $\mu_\alpha$, $\mu_\delta$, and $\dot{f}_{\rm s, obs}$, we used the values given in version 1.70 of the ATNF pulsar catalogue \citep{mhth05} for those 237 pulsars. For $d$ and $f_{\rm s}$, we used the values given in Tables 3 and 4 of \citet{aa21}. For five of those pulsars, distance values are not available and so $h_{0}^{\rm sd}$ values are not reported for these pulsars in \citet{aa21}. We dropped these five pulsars from our calculations as well and calculated $\left( \frac{\dot{P}_{\rm s} }{ {P}_{\rm s}} \right)_{\rm excess}$ values for the remaining pulsars using GalDynPsr.

Moreover, \citet{aa21} found 13 pulsars (out of the remaining 232 pulsars) for which, after correcting for the dynamical effects, the $\dot{P}_{\rm s}$ values turned out to be negative ($\dot{P}_{\rm s, int}$ should be positive) and so $h_{0}^{\rm sd}$ values are not reported by \citet{aa21} for these pulsars either. However, for three (PSRs J1125+7819, J1300+1240, and J1832-0836) out of these 13 pulsars, we got a positive $\dot{P}_{\rm s,int}$ value using GalDynPsr and so we calculated the $h_{0}^{\rm sd}$ value for these three. For the remaining ten pulsars (out of 13), we also got a negative value of $\dot{P}_{\rm s,int}$ and we dropped those pulsars from our calculation of $h_{0}^{\rm sd}$. After using GalDynPsr we found an additional five pulsars with negative $\dot{P}_{\rm s, int}$ values and dropped these pulsars too from our calculation of $h_{0}^{\rm sd}$.

For 77 of the 232 pulsars, proper motion values are not reported in the ATNF pulsar catalogue and so, for those pulsars, $\left( \frac{\dot{P}_{\rm s} }{ {P}_{\rm s}} \right)_{\rm excess}$ does not have a contribution from $\left( \frac{\dot{P}_{\rm s} }{ {P}_{\rm s}} \right)_{\rm excess,Shk}$. In the future, with improvements in timing precision, the proper motion values might be measured. Additionally, eight out of the 232 pulsars are globular cluster pulsars. For these pulsars, there will be an additional contribution to $\left( \frac{\dot{P}_{\rm s} }{ {P}_{\rm s}} \right)_{\rm excess}$ from the respective cluster potentials. Further investigation is required in order to calculate the respective cluster potentials and accelerations resulting from them, and hence, for both sets of pulsars discussed above (for which a complete contribution to $\left( \frac{\dot{P}_{\rm s} }{ {P}_{\rm s}} \right)_{\rm excess}$ is unavailable), after subtracting the available dynamical effects from the $\dot{f}_{\rm s,obs}$ values we actually got residual frequency derivative values ($\dot{f}_{\rm s,res}$) instead of truly intrinsic frequency derivative values.

Finally, after ignoring the cases without distance measurements, with negative $\dot{P}_{\rm s,int}$ values, without proper motion measurements, and those in globular clusters, we are left with 139 pulsars. Table \ref{tb:horiall} contains the parameters for all these 139 pulsars. For five of these pulsars (indicated by superscript $\epsilon$), \citet{aa21} did not correct for dynamical effects. For 93 of the 139 pulsars, there is an increase in the $\dot{f}_{\rm s,int}$ values, implying an increase in $h_{0}^{\rm sd}$ values when we compare the values calculated by us with those reported by \citet{aa21}. There are two pulsars (PSRs J2222-0137 and J1400-1431) for which the percentage increase in $h_{0}^{\rm sd}$ values is greater than $100\%$. For three pulsars (PSRs J2322-2650, J1709+2313, and J2010-1323), the percentage increase in $h_{0}^{\rm sd}$ values is between $20\%$ and $100\%$. The first three rows of table \ref{tb:horiall} show the pulsars for which \citet{aa21} got a negative $\dot{P}_{\rm s,int}$ value and did not calculate $h_{0}^{\rm sd}$, whereas we got a positive $\dot{P}_{\rm s,int}$ value and calculated $h_{0}^{\rm sd}$.

We note that \citet{aa21} do not take into account errors in their estimation of the spin-down limits. We also do not provide error estimates as the Milky Way potential model that we use from galpy does not return errors in their potential and acceleration values. However, in the future, once we have a Galactic potential model that returns error values, it will not be a difficult task to incorporate that into GalDynPsr.

\section{Conclusion}

We know that for a pulsar to be a good candidate for the detection of continuous gravitational waves, that is, for it to be a high-value target, the ratio $\frac{h_{0}^{95\%}}{h_{0}^{\rm sd}}$ should be less than one, where $h_{0}^{95\%}$ is the 95\% credible upper limit of the gravitational wave amplitude, $h_0$. Along with a precise measurement of $h_{0}^{95\%}$, an accurate estimate of $h_{0}^{\rm sd}$ will improve the estimate of the ratio. 

As mentioned previously, for calculating the $h_{0}^{\rm sd}$ values using Eq. (\ref{eq:h0sd}), $\dot{f}_{\rm s,int}$ values devoid of contamination from dynamical effects should be used. We used GalDynPsr to calculate the $\dot{f}_{\rm s,int}$ values for the pulsars given in \citet{aa21}. 

We found that for two pulsars (PSRs J2222-0137 and J1400-1431) the percentage increase in $h_{0}^{\rm sd}$ values was greater than $100\%$ and for three pulsars (PSRs J2322-2650, J1709+2313, and J2010-1323) the percentage increase in $h_{0}^{\rm sd}$ values was between $20\%$ and $100\%$. We also got a positive $\dot{P}_{\rm s,int}$ value for three pulsars (PSRs J1125+7819, J1300+1240, and J1832-0836) using GalDynPsr, for which \citet{aa21} got a negative value of $\dot{P}_{\rm s,int}$. We calculated the $h_{0}^{\rm sd}$ value for these pulsars too. 

We suggest using Model Lb in GalDynPsr to calculate a more realistic value of $\dot{f}_{\rm s,int}$ before using it in the estimation of $h_{0}^{\rm sd}$ (given the availability of all the required observables for estimating $\dot{f}_{\rm s,int}$ using GalDynPsr). For the pulsars present inside globular clusters, there will be an additional contribution from the respective cluster potentials to the $\left( \frac{\dot{P}_{\rm s} }{ {P}_{\rm s}} \right)_{\rm excess}$ term, and hence further investigation of cluster potential is required in order to calculate the true intrinsic value of the spin frequency derivative.

We know that pulsar distances are estimated through various methods, including from VLBI \citep{dwnc13,dwnc18} or timing parallaxes \citep{bas16,dcl16}, from dispersion measure values using a model of free electron distribution in the Galaxy (such as the NE2001 model \citep{cl02,cl03}), or from objects of known distance with which the pulsar has an association (such as globular clusters \citep{harris96}). However, some methods are more accurate than others. Distances estimated from parallax values are generally more accurate than those measured using dispersion measure values. For example, the distance estimates for pulsars PSR J0437-4715 and PSR J1909-3744 are fairly accurate (based on timing parallaxes) and, for them, the percentage change in the $h_{0}^{\rm sd}$ values was found to be -4.41\% and 0.24\%, respectively.
In the future, with the increase in the number of parallax measurements of pulsars, their distance estimates will improve and using those estimates along with our formalism will give more realistic values of $h_{0}^{\rm sd}$ for those pulsars.

As a future scope of work, our method could turn out to be important for future gravitational wave searches. Using GalDynPsr, we got more realistic values of $\dot{f}_{\rm s,int}$, which will improve the phase of the signal, as the phase evolution of the signal is dependent on the spin frequency and its derivatives. This may enhance the sensitivity of search pipelines for the continuous gravitational wave signals.

\begin{acknowledgements} 
We are grateful to Prof. Wynn Ho for his suggestions and for drawing our attention to the recent work of \citet{aa21}. We would also like to thank the LIGO-Continuous Wave group, particularly Prof. Matthew Bailes, for their useful comments. We also thank Prof. Manjari Bagchi for her contribution to the formulation of GalDynPsr and for envisioning its applications. We are also grateful to Amy Hewitt and Cristiano Palomba for the information regarding the method used in \citet{aa21} for the correction of dynamical effects.
\end{acknowledgements}
\bibliographystyle{aa}
\bibliography{ref.bib}

\begin{appendix}

    \section{Table}

\onecolumn
\begin{landscape}

\LTcapwidth=1.0\linewidth
\begin{longtable}{@{\hskip1.pt}l@{\hskip1.pt}@{\hskip1.pt}l@{\hskip1.pt}@{\hskip1.pt}l@{\hskip1.pt}@{\hskip1.pt}l@{\hskip1.pt}@{\hskip1.pt}l@{\hskip1.pt}@{\hskip1.pt}l@{\hskip1.pt}@{\hskip1.pt}l@{\hskip1.pt}@{\hskip1.pt}l@{\hskip1.pt}@{\hskip1.pt}l@{\hskip1.pt}@{\hskip1.pt}l@{\hskip1.pt}@{\hskip1.pt}l@{\hskip1.pt}@{\hskip1.pt}l@{\hskip1.pt}@{\hskip1.pt}l@{\hskip1.pt}@{\hskip1.pt}l@{\hskip1.pt}@{\hskip1.pt}l@{\hskip1.pt}}
\caption{Parameters of the 139 pulsars for which all the required observables are available to calculate the complete dynamical effect contribution, using GalDynPsr to calculate the $\dot{f}_{\rm s, int}$ values. Here, $l$ is the Galactic longitude, $b$ is the Galactic latitude, $d$ is the distance between the pulsar and the Solar System barycentre, $\mu_{\alpha}$ is the proper motion in the direction of the right ascension, $\mu_{\delta}$ is the proper motion in the declination, $f_{\rm s}$ is the spin frequency, $\dot{f}_{\rm s, obs}$ is the observed value of the first derivative of the spin frequency, $\dot{f}_{\rm s, int}$ is the intrinsic value of the first derivative of the spin frequency, $h_{0}^{\rm sd}$ is the spin-down limit of the pulsar (assumed to be a rotating triaxial star) calculated in this work, $h_{0}^{95\%}$ is the 95\% credible upper limit of the gravitational wave amplitude $h_0$, and $h_{0,AA}^{\rm sd}$ is the spin-down limit of the pulsar (assumed to be a rotating triaxial star) as given in \citet{aa21}. The last column shows the percentage increase in the spin-down limit values. We display the results till the third decimal place. $\dagger$: Pulsars for which \citet{aa21} got a positive $\dot{f}_{\rm s,int}$ but we got a negative $\dot{f}_{\rm s,int}$. $\epsilon$: Pulsars for which the spin frequency derivative was not corrected for dynamical effects by \citet{aa21}.} \label{tb:horiall} \\
 \hline 
Pulsars & $l$  & $b$  & $d$  & $\mu_{\alpha}$  &  $\mu_{\delta}$ & $f_{\rm s}$  & $\dot{f}_{\rm s,obs}$  &$\left(\frac{\dot{f}}{f}\right)_{\rm s,excess}$ & $\dot{f}_{\rm s,int}$  & $h_{0}^{95\%}$ & $h_{0}^{\rm sd}$  & \addstackgap{$\frac{h_{0}^{95\%}}{h_{0}^{\rm sd}}$} & $h_{0,AA}^{\rm sd}$  & \addstackgap{$\frac{(h_{0}^{\rm sd}-h_{0,AA}^{\rm sd}) \times 100}{h_{0,AA}^{\rm sd}}$} \\

& (deg)  & (deg)  & (kpc)  & (mas/yr)  & (mas/yr)  & (Hz)  & ($\times10^{-14}~\rm s^{-2}$)  & ($\times10^{-19}~\rm s^{-1}$)  & ($\times10^{-14}~\rm s^{-2}$)  & ($\times10^{-26}$)  & ($\times10^{-27}$)  &   & ($\times10^{-27}$) &   \\

\endfirsthead

\multicolumn{7}{c}%
{{\bfseries (Table \ref{tb:horiall} continued)}} \\
\hline  

Pulsars & $l$  & $b$  & $d$  & $\mu_{\alpha}$  &  $\mu_{\delta}$ & $f_{\rm s}$  & $\dot{f}_{\rm s,obs}$  &$\left(\frac{\dot{f}}{f}\right)_{\rm s,excess}$ & $\dot{f}_{\rm s,int}$  & $h_{0}^{95\%}$ & $h_{0}^{\rm sd}$  & \addstackgap{$\frac{h_{0}^{95\%}}{h_{0}^{\rm sd}}$} & $h_{0,AA}^{\rm sd}$  & \addstackgap{$\frac{(h_{0}^{\rm sd}-h_{0,AA}^{\rm sd}) \times 100}{h_{0,AA}^{\rm sd}}$} \\

& (deg)  & (deg)  & (kpc)  & (mas/yr)  & (mas/yr)  & (Hz)  & ($\times10^{-14}~\rm s^{-2}$)  & ($\times10^{-19}~\rm s^{-1}$)  & ($\times10^{-14}~\rm s^{-2}$)  & ($\times10^{-26}$)  & ($\times10^{-27}$)  &   & ($\times10^{-27}$) &   \\ \hline 
\endhead

\hline
J1125+7819$^\dagger$  &  128.289  &  37.895  &  0.88  &  27.34  &  -2.07  &  238.0  &  -0.039  &  -15.213  &  -0.003  &  0.726  &  0.332  &  21.868  &  0.0  &  $-$ \\  
 \hline 
J1300+1240$^\dagger$  &  311.31  &  75.414  &  0.6  &  45.5  &  -84.7  &  160.8  &  -0.296  &  -132.989  &  -0.082  &  1.1  &  3.032  &  3.628  &  0.0  &  $-$ \\  
 \hline 
J1832-0836$^\dagger$  &  23.109  &  0.257  &  1.6  &  -7.99  &  -21.46  &  367.8  &  -0.112  &  -21.611  &  -0.032  &  0.83  &  0.473  &  17.539  &  0.0  &  $-$ \\  
 \hline 
J2222-0137  &  62.018  &  -46.075  &  0.27  &  44.7  &  -5.69  &  30.5  &  -0.054  &  -12.516  &  -0.05  &  1.39  &  12.103  &  1.148  &  2.05  &  490.404 \\  
 \hline 
J1400-1431  &  326.986  &  45.087  &  0.27  &  17.0  &  -55.0  &  324.2  &  -0.076  &  -21.048  &  -0.008  &  0.831  &  1.465  &  5.671  &  0.51  &  187.342 \\  
 \hline 
J2322-2650  &  28.637  &  -70.228  &  0.23  &  -2.4  &  -8.3  &  288.8  &  -0.005  &  0.606  &  -0.007  &  1.13  &  1.678  &  6.734  &  1.21  &  38.675 \\  
 \hline 
J1709+2313  &  44.522  &  32.209  &  2.18  &  -3.2  &  -9.7  &  215.9  &  -0.017  &  -3.359  &  -0.01  &  0.809  &  0.248  &  32.68  &  0.193  &  28.264 \\  
 \hline 
J2010-1323  &  29.446  &  -23.54  &  1.16  &  2.55  &  -5.86  &  191.5  &  -0.018  &  -0.681  &  -0.016  &  1.44  &  0.643  &  22.394  &  0.529  &  21.557 \\  
 \hline 
J1640+2224  &  41.051  &  38.271  &  1.52  &  2.078  &  -11.336  &  316.1  &  -0.028  &  -3.115  &  -0.018  &  0.745  &  0.404  &  18.45  &  0.337  &  19.821 \\  
 \hline 
J1730-2304  &  3.137  &  6.023  &  0.47  &  20.12  &  -3.0  &  123.1  &  -0.031  &  -5.168  &  -0.024  &  0.5  &  2.407  &  2.077  &  2.03  &  18.589 \\  
 \hline 
J0610-2100  &  227.747  &  -18.184  &  3.26  &  9.11  &  16.45  &  259.0  &  -0.083  &  -28.261  &  -0.009  &  0.939  &  0.15  &  62.746  &  0.132  &  13.372 \\  
 \hline 
J1518+4904  &  80.808  &  54.282  &  0.96  &  -0.67  &  -8.53  &  24.4  &  -0.002  &  0.137  &  -0.002  &  0.943  &  0.692  &  13.625  &  0.629  &  10.029 \\  
 \hline 
J0154+1833  &  143.185  &  -41.81  &  1.62  &  10.3  &  -8.9  &  422.9  &  -0.052  &  -6.765  &  -0.024  &  0.974  &  0.372  &  26.197  &  0.339  &  9.677 \\  
 \hline 
J1747-4036  &  350.208  &  -6.412  &  7.15  &  -1.41  &  -2.3  &  607.7  &  -0.485  &  0.16  &  -0.486  &  1.12  &  0.319  &  35.103  &  0.292  &  9.269 \\  
 \hline 
J1411+2551  &  33.38  &  72.101  &  1.13  &  -3.0  &  -4.0  &  16.0  &  -0.002  &  1.55  &  -0.003  &  3.18  &  0.927  &  34.312  &  0.853  &  8.65 \\  
 \hline 
J1453+1902  &  23.395  &  60.812  &  1.27  &  0.9  &  -10.9  &  172.6  &  -0.035  &  -1.483  &  -0.032  &  0.804  &  0.867  &  9.272  &  0.799  &  8.522 \\  
 \hline 
J1012+5307  &  160.347  &  50.858  &  0.7  &  2.64  &  -25.54  &  190.3  &  -0.062  &  -10.359  &  -0.042  &  0.78  &  1.717  &  4.542  &  1.62  &  6.016 \\  
 \hline 
J2229+2643  &  87.693  &  -26.284  &  1.8  &  -1.89  &  -5.76  &  335.8  &  -0.017  &  0.327  &  -0.018  &  0.668  &  0.331  &  20.207  &  0.312  &  5.954 \\  
 \hline  
J2256-1024  &  59.231  &  -58.293  &  1.33  &  3.2  &  -8.5  &  435.8  &  -0.216  &  -0.399  &  -0.214  &  1.05  &  1.343  &  7.816  &  1.27  &  5.781 \\  
 \hline 
J1312+0051  &  314.842  &  63.227  &  1.47  &  -22.35  &  -11.2  &  236.5  &  -0.098  &  -19.873  &  -0.051  &  0.629  &  0.806  &  7.804  &  0.762  &  5.778 \\  
 \hline 
J2322+2057  &  96.515  &  -37.31  &  1.01  &  -18.4  &  -15.4  &  208.0  &  -0.042  &  -12.655  &  -0.015  &  0.625  &  0.688  &  9.08  &  0.654  &  5.25 \\  
 \hline 
J0645+5158  &  163.963  &  20.251  &  1.2  &  1.546  &  -7.47  &  112.9  &  -0.006  &  -2.185  &  -0.004  &  0.663  &  0.39  &  16.98  &  0.371  &  5.247 \\  
 \hline 
J0101-6422  &  301.192  &  -52.72  &  1.0  &  10.0  &  -12.0  &  388.6  &  -0.078  &  -4.028  &  -0.062  &  0.74  &  1.021  &  7.25  &  0.971  &  5.119 \\  
 \hline 
J2317+1439  &  91.361  &  -42.36  &  2.16  &  -1.305  &  3.45  &  290.3  &  -0.02  &  1.543  &  -0.025  &  1.09  &  0.346  &  31.486  &  0.33  &  4.906 \\  
 \hline 
J1933-6211  &  334.431  &  -28.632  &  0.65  &  -5.54  &  10.7  &  282.2  &  -0.031  &  -1.806  &  -0.026  &  0.619  &  1.185  &  5.225  &  1.13  &  4.846 \\  
 \hline 
J1022+1001  &  231.795  &  51.101  &  0.64  &  -16.0  &  2.0  &  60.8  &  -0.016  &  -2.903  &  -0.014  &  0.77  &  1.929  &  3.992  &  1.84  &  4.828 \\  
 \hline 
J1421-4409  &  319.497  &  15.809  &  2.08  &  -11.6  &  -7.9  &  156.6  &  -0.03  &  -9.099  &  -0.016  &  0.662  &  0.393  &  16.851  &  0.376  &  4.483 \\  
 \hline 
J0613-0200  &  210.413  &  -9.305  &  0.6  &  1.834  &  -10.353  &  326.6  &  -0.102  &  -1.842  &  -0.096  &  1.64  &  2.308  &  7.106  &  2.21  &  4.428 \\  
 \hline 
J1745+1017  &  34.869  &  19.254  &  1.21  &  6.0  &  -5.0  &  377.1  &  -0.039  &  -1.393  &  -0.034  &  0.829  &  0.629  &  13.187  &  0.602  &  4.423 \\  
 \hline 
J1713+0747  &  28.751  &  25.223  &  1.0  &  4.924  &  -3.915  &  218.8  &  -0.041  &  -0.46  &  -0.04  &  0.688  &  1.088  &  6.323  &  1.05  &  3.634 \\  
 \hline 
J1614-2230  &  352.636  &  20.192  &  0.7  &  3.86  &  -32.2  &  317.4  &  -0.097  &  -17.975  &  -0.04  &  0.991  &  1.292  &  7.673  &  1.25  &  3.329 \\  
 \hline 
J2235+1506  &  80.878  &  -36.444  &  1.54  &  15.0  &  10.0  &  16.7  &  -0.004  &  -10.13  &  -0.003  &  1.92  &  0.668  &  28.726  &  0.65  &  2.827 \\  
 \hline 
J0034-0534  &  111.492  &  -68.069  &  1.35  &  7.9  &  -9.2  &  532.7  &  -0.141  &  -2.897  &  -0.125  &  1.06  &  0.917  &  11.56  &  0.892  &  2.795 \\  
 \hline 
J2051-0827  &  39.192  &  -30.411  &  1.47  &  5.63  &  2.34  &  221.8  &  -0.063  &  -0.01  &  -0.063  &  0.57  &  0.922  &  6.183  &  0.899  &  2.537 \\  
 \hline 
J1622-6617  &  321.978  &  -11.56  &  4.05  &  -3.0  &  6.0  &  42.3  &  -0.011  &  -2.325  &  -0.01  &  0.574  &  0.299  &  19.178  &  0.292  &  2.501 \\  
 \hline 
J1903-7051  &  324.391  &  -26.508  &  0.93  &  -8.8  &  -16.0  &  277.9  &  -0.081  &  -6.843  &  -0.062  &  1.06  &  1.291  &  8.211  &  1.26  &  2.453 \\  
 \hline 
J2033+1734  &  60.857  &  -13.154  &  1.74  &  -5.8  &  -9.7  &  168.1  &  -0.031  &  -3.961  &  -0.025  &  0.675  &  0.563  &  11.981  &  0.551  &  2.25 \\  
 \hline 
J2019+2425  &  64.746  &  -6.624  &  1.16  &  -9.41  &  -20.6  &  254.2  &  -0.045  &  -13.645  &  -0.011  &  1.06  &  0.451  &  23.515  &  0.441  &  2.217 \\  
 \hline 
J2055+3829  &  80.615  &  -4.259  &  4.59  &  5.92  &  0.79  &  478.6  &  -0.023  &  -0.546  &  -0.02  &  1.19  &  0.114  &  104.031  &  0.112  &  2.133 \\  
 \hline 
J0023+0923  &  111.383  &  -52.849  &  1.11  &  -12.44  &  -6.16  &  327.8  &  -0.123  &  -3.626  &  -0.111  &  1.74  &  1.336  &  13.02  &  1.31  &  2.013 \\  
 \hline 
J1231-1411  &  295.531  &  48.385  &  0.42  &  -62.03  &  6.2  &  271.5  &  -0.168  &  -38.501  &  -0.063  &  1.06  &  2.936  &  3.611  &  2.88  &  1.94 \\  
 \hline 
J1630+3734  &  60.245  &  43.215  &  1.19  &  2.4  &  -15.9  &  301.4  &  -0.098  &  -5.587  &  -0.081  &  0.863  &  1.111  &  7.77  &  1.09  &  1.903 \\  
 \hline 
J1455-3330  &  330.722  &  22.562  &  1.01  &  7.88  &  -2.07  &  125.2  &  -0.038  &  -1.26  &  -0.037  &  0.545  &  1.364  &  3.996  &  1.34  &  1.771 \\  
 \hline 
J1412+7922$^\epsilon$  &  118.318  &  37.02  &  2.0  &  -40.0  &  -56.0  &  16.9  &  -97.172  &  -228.943  &  -97.133  &  2.47  &  96.672  &  0.256  &  95.1  &  1.653 \\  
 \hline 
J1731-1847  &  6.89  &  8.151  &  4.78  &  -1.7  &  -6.0  &  426.5  &  -0.462  &  -9.189  &  -0.423  &  0.865  &  0.531  &  16.279  &  0.523  &  1.596 \\  
 \hline 
J1337-6423  &  307.889  &  -1.958  &  5.94  &  -6.0  &  -7.0  &  106.1  &  -0.028  &  -6.845  &  -0.021  &  0.668  &  0.189  &  35.363  &  0.186  &  1.557 \\  
 \hline 
J0534+2200  &  184.558  &  -5.784  &  2.0  &  -14.7  &  2.0  &  29.6  &  -37753.5  &  -12.186  &  -37753.496  &  1.34  &  1440.104  &  0.009  &  1420.0  &  1.416 \\  
 \hline 
J1902-5105  &  345.65  &  -22.379  &  1.65  &  -4.8  &  -4.4  &  573.9  &  -0.303  &  -1.703  &  -0.293  &  1.42  &  1.105  &  12.853  &  1.09  &  1.361 \\  
 \hline 
J0931-1902  &  250.999  &  23.054  &  3.72  &  -2.47  &  -4.31  &  215.6  &  -0.017  &  -0.853  &  -0.015  &  0.771  &  0.181  &  42.597  &  0.179  &  1.117 \\  
 \hline 
J1207-5050  &  295.861  &  11.422  &  1.27  &  6.9  &  1.4  &  206.5  &  -0.026  &  -0.493  &  -0.025  &  1.1  &  0.696  &  15.794  &  0.689  &  1.087 \\  
 \hline 
J1944+0907  &  47.16  &  -7.357  &  1.22  &  14.01  &  -22.59  &  192.9  &  -0.064  &  -20.652  &  -0.025  &  0.617  &  0.747  &  8.258  &  0.74  &  0.969 \\  
 \hline 
J2302+4442  &  103.395  &  -14.005  &  0.86  &  0.18  &  -5.8  &  192.6  &  -0.051  &  0.051  &  -0.052  &  0.779  &  1.534  &  5.078  &  1.52  &  0.92 \\  
 \hline 
J1431-4715  &  320.051  &  12.253  &  1.53  &  -12.01  &  -14.51  &  497.0  &  -0.349  &  -12.878  &  -0.285  &  1.57  &  1.261  &  12.447  &  1.25  &  0.909 \\  
 \hline 
J2017+0603  &  48.621  &  -16.026  &  1.4  &  2.21  &  0.15  &  345.3  &  -0.095  &  0.673  &  -0.098  &  0.77  &  0.969  &  7.95  &  0.96  &  0.89 \\  
 \hline 
J2241-5236  &  337.457  &  -54.927  &  1.05  &  18.881  &  -5.294  &  457.3  &  -0.144  &  -7.96  &  -0.108  &  0.881  &  1.179  &  7.47  &  1.17  &  0.805 \\  
 \hline 
J1738+0333  &  27.721  &  17.742  &  1.47  &  7.037  &  5.073  &  170.9  &  -0.07  &  -2.608  &  -0.066  &  0.484  &  1.078  &  4.489  &  1.07  &  0.776 \\  
 \hline 
J1719-1438  &  8.858  &  12.838  &  0.34  &  1.9  &  -11.0  &  172.7  &  -0.024  &  -1.23  &  -0.022  &  0.719  &  2.669  &  2.694  &  2.65  &  0.719 \\  
 \hline 
J1955+2527  &  62.74  &  -1.571  &  8.18  &  -1.9  &  -2.4  &  205.2  &  -0.038  &  4.432  &  -0.047  &  0.978  &  0.15  &  65.204  &  0.149  &  0.665 \\  
 \hline 
J2043+1711  &  61.919  &  -15.313  &  1.6  &  -5.703  &  -10.841  &  420.2  &  -0.093  &  -4.409  &  -0.074  &  1.07  &  0.669  &  15.989  &  0.665  &  0.632 \\  
 \hline 
J0751+1807  &  202.73  &  21.086  &  0.6  &  -2.73  &  -13.4  &  287.5  &  -0.064  &  -2.785  &  -0.056  &  1.01  &  1.882  &  5.368  &  1.87  &  0.623 \\  
 \hline 
J0348+0432  &  183.337  &  -36.774  &  2.1  &  4.04  &  3.5  &  25.6  &  -0.016  &  -1.951  &  -0.015  &  1.38  &  0.937  &  14.733  &  0.931  &  0.611 \\  
 \hline 
J0453+1559  &  184.125  &  -17.137  &  0.52  &  -5.5  &  -6.0  &  21.8  &  -0.009  &  -1.079  &  -0.009  &  1.1  &  3.088  &  3.562  &  3.07  &  0.6 \\  
 \hline 
J0636+5129  &  163.908567  &  18.642582  &  0.21  &  3.22  &  -1.61  &  348.6  &  -0.042  &  -0.106  &  -0.042  &  0.784  &  4.192  &  1.87  &  4.17  &  0.523 \\  
 \hline 
J1911+1347  &  47.518  &  1.809  &  1.36  &  -2.944  &  -3.71  &  216.2  &  -0.079  &  -0.534  &  -0.078  &  0.565  &  1.126  &  5.018  &  1.12  &  0.523 \\  
 \hline 
J0509+0856  &  192.486  &  -17.931  &  0.82  &  5.4  &  -4.3  &  246.6  &  -0.027  &  -1.31  &  -0.024  &  0.828  &  0.962  &  8.608  &  0.957  &  0.513 \\  
 \hline 
J1903+0327  &  37.336  &  -1.014  &  6.12  &  -2.06  &  -5.21  &  465.1  &  -0.407  &  0.007  &  -0.407  &  1.04  &  0.39  &  26.672  &  0.388  &  0.497 \\  
 \hline 
J1446-4701  &  322.5  &  11.425  &  1.5  &  -4.37  &  -3.02  &  455.6  &  -0.204  &  -0.906  &  -0.199  &  1.2  &  1.125  &  10.666  &  1.12  &  0.449 \\  
 \hline 
J1955+2908  &  65.839  &  0.443  &  6.3  &  -0.98  &  -4.05  &  163.0  &  -0.079  &  2.563  &  -0.083  &  0.552  &  0.289  &  19.082  &  0.288  &  0.445 \\  
 \hline 
J0030+0451  &  113.141  &  -57.611  &  0.33  &  -6.33  &  0.8  &  205.5  &  -0.043  &  0.741  &  -0.044  &  0.654  &  3.596  &  1.819  &  3.58  &  0.445 \\  
 \hline 
J1502-6752  &  314.799  &  -8.067  &  7.73  &  -6.0  &  -14.0  &  37.4  &  -0.044  &  -35.822  &  -0.031  &  0.701  &  0.298  &  23.5  &  0.297  &  0.435 \\  
 \hline 
J0709+0458  &  210.498  &  6.2  &  1.2  &  -0.7  &  -1.3  &  29.0  &  -0.032  &  -0.615  &  -0.032  &  1.26  &  2.229  &  5.652  &  2.22  &  0.426 \\  
 \hline 
J1911-1114  &  25.137  &  -9.579  &  1.07  &  -13.75  &  -9.1  &  275.8  &  -0.106  &  -7.537  &  -0.085  &  0.639  &  1.326  &  4.82  &  1.32  &  0.425 \\  
 \hline 
J1952+3252  &  68.765  &  2.823  &  3.0  &  -28.8  &  -14.7  &  25.3  &  -374.016  &  -73.689  &  -373.997  &  1.79  &  103.358  &  0.173  &  103.0  &  0.348 \\  
 \hline 
J1125-5825  &  291.893  &  2.602  &  1.74  &  -10.0  &  2.4  &  322.4  &  -0.633  &  -3.171  &  -0.623  &  0.919  &  2.037  &  4.512  &  2.03  &  0.333 \\  
 \hline 
J1939+2134  &  57.509  &  -0.29  &  4.8  &  0.072  &  -0.407  &  641.9  &  -4.331  &  4.086  &  -4.357  &  2.2  &  1.384  &  15.893  &  1.38  &  0.309 \\  
 \hline 
J1537+1155  &  19.848  &  48.341  &  1.05  &  1.482  &  -25.285  &  26.4  &  -0.169  &  -14.773  &  -0.165  &  1.18  &  6.067  &  1.945  &  6.05  &  0.278 \\  
 \hline 
J1950+2414  &  61.098  &  -1.169  &  7.27  &  -2.12  &  -3.64  &  232.3  &  -0.101  &  2.986  &  -0.108  &  0.909  &  0.24  &  37.931  &  0.239  &  0.271 \\  
 \hline 
J1909-3744  &  359.731  &  -19.596  &  1.15  &  -9.514  &  -35.777  &  339.3  &  -0.161  &  -38.584  &  -0.031  &  1.17  &  0.666  &  17.578  &  0.664  &  0.244 \\  
 \hline 
J1035-6720  &  290.371  &  -7.843  &  1.46  &  -12.0  &  1.0  &  348.2  &  -0.563  &  -3.932  &  -0.55  &  1.27  &  2.195  &  5.787  &  2.19  &  0.209 \\  
 \hline 
J1949+3106  &  66.858  &  2.554  &  7.47  &  -2.894  &  -5.093  &  76.1  &  -0.054  &  -0.58  &  -0.054  &  0.779  &  0.288  &  27.094  &  0.287  &  0.179 \\  
 \hline 
J0614-3329  &  240.501  &  -21.827  &  0.63  &  0.58  &  -1.92  &  317.6  &  -0.177  &  0.471  &  -0.178  &  0.831  &  3.035  &  2.738  &  3.03  &  0.158 \\  
 \hline 
J0900-3144  &  256.162  &  9.486  &  0.89  &  -1.01  &  2.02  &  90.0  &  -0.04  &  0.563  &  -0.04  &  0.624  &  1.913  &  3.262  &  1.91  &  0.155 \\  
 \hline 
J1732-5049  &  340.029  &  -9.454  &  1.88  &  -0.5  &  -9.87  &  188.2  &  -0.05  &  -5.561  &  -0.04  &  0.668  &  0.624  &  10.706  &  0.623  &  0.154 \\  
 \hline 
J0340+4130  &  153.783  &  -11.022  &  1.6  &  -0.53  &  -3.3  &  303.1  &  -0.065  &  -1.198  &  -0.061  &  0.68  &  0.716  &  9.501  &  0.715  &  0.096 \\  
 \hline 
J1833-0827  &  23.386  &  0.063  &  4.5  &  -31.0  &  13.0  &  11.7  &  -126.16  &  -126.011  &  -126.145  &  9.25  &  58.847  &  1.572  &  58.8  &  0.079 \\  
 \hline 
J1302-6350  &  304.184  &  -0.992  &  2.3  &  -7.01  &  -0.53  &  20.9  &  -99.89  &  -1.503  &  -99.89  &  2.42  &  76.657  &  0.316  &  76.6  &  0.074 \\  
 \hline 
J1514-4946  &  325.249  &  6.807  &  0.91  &  -0.3  &  -30.0  &  278.6  &  -0.145  &  -20.123  &  -0.089  &  1.04  &  1.581  &  6.578  &  1.58  &  0.066 \\  
 \hline 
J2214+3000  &  86.855  &  -21.665  &  0.6  &  20.92  &  -1.94  &  320.6  &  -0.151  &  -5.609  &  -0.133  &  0.832  &  2.742  &  3.035  &  2.74  &  0.065 \\  
 \hline 
J1915+1606  &  49.968  &  2.122  &  5.25  &  -0.72  &  -0.03  &  16.9  &  -0.247  &  4.352  &  -0.248  &  3.3  &  1.861  &  17.731  &  1.86  &  0.06 \\  
 \hline 
J1923+2515  &  58.946  &  4.749  &  1.2  &  -6.92  &  -14.28  &  264.0  &  -0.067  &  -6.705  &  -0.049  &  0.65  &  0.915  &  7.108  &  0.914  &  0.055 \\  
 \hline 
J1600-3053  &  344.09  &  16.451  &  3.0  &  -0.961  &  -6.97  &  277.9  &  -0.073  &  -4.357  &  -0.061  &  0.83  &  0.399  &  20.792  &  0.399  &  0.046 \\  
 \hline 
J1741+1351  &  37.885  &  21.641  &  1.08  &  -8.98  &  -7.41  &  266.9  &  -0.215  &  -2.981  &  -0.207  &  0.662  &  2.081  &  3.181  &  2.08  &  0.039 \\  
 \hline 
J1756-2251  &  6.499  &  0.948  &  0.73  &  -2.42  &  0.0  &  35.1  &  -0.126  &  -0.869  &  -0.125  &  0.613  &  6.601  &  0.929  &  6.6  &  0.012 \\  
 \hline 
J1959+2048  &  59.197  &  -4.697  &  1.4  &  -16.0  &  -25.8  &  622.1  &  -0.652  &  -30.56  &  -0.462  &  1.31  &  1.57  &  8.344  &  1.57  &  0.0 \\  
 \hline 
J1918-0642  &  30.027  &  -9.123  &  1.1  &  -7.149  &  -5.97  &  130.8  &  -0.044  &  -2.654  &  -0.04  &  0.997  &  1.29  &  7.73  &  1.29  &  -0.022 \\  
 \hline 
J2053+4650  &  86.861  &  1.302  &  3.81  &  -2.8  &  -5.4  &  79.5  &  -0.109  &  -0.753  &  -0.108  &  0.527  &  0.781  &  6.75  &  0.781  &  -0.035 \\  
 \hline 
J2045+3633  &  77.832  &  -3.926  &  5.63  &  -3.2  &  -2.68  &  31.6  &  -0.059  &  1.665  &  -0.059  &  0.654  &  0.62  &  10.552  &  0.62  &  -0.037 \\  
 \hline 
J1813-1749  &  12.816  &  -0.02  &  6.2  &  -5.0  &  -13.2  &  22.4  &  -6344.5  &  -35.645  &  -6344.492  &  1.53  &  218.914  &  0.07  &  219.0  &  -0.039 \\  
 \hline 
J1853+1303  &  44.875  &  5.367  &  1.32  &  -1.62  &  -2.9  &  244.4  &  -0.052  &  -0.201  &  -0.052  &  1.01  &  0.888  &  11.38  &  0.888  &  -0.053 \\  
 \hline 
J0218+4232  &  139.508  &  -17.527  &  3.15  &  5.35  &  -3.74  &  430.5  &  -1.434  &  -3.929  &  -1.417  &  0.894  &  1.469  &  6.086  &  1.47  &  -0.071 \\  
 \hline 
J0908-4913$^\epsilon$  &  270.265  &  -1.019  &  1.0  &  -37.0  &  31.0  &  9.4  &  -132.482  &  -55.795  &  -132.477  &  27.6  &  302.759  &  0.912  &  303.0  &  -0.079 \\  
 \hline 
J1809-1917  &  11.094  &  0.08  &  3.27  &  -19.0  &  50.0  &  12.1  &  -372.788  &  -231.624  &  -372.76  &  9.0  &  136.888  &  0.657  &  137.0  &  -0.082 \\  
 \hline 
J1545-4550$^\epsilon$  &  331.892  &  6.988  &  2.2  &  -0.48  &  2.37  &  279.7  &  -0.41  &  -1.158  &  -0.407  &  0.669  &  1.399  &  4.783  &  1.4  &  -0.103 \\  
 \hline 
J0835-4510  &  263.552  &  -2.787  &  0.28  &  -49.68  &  29.9  &  11.2  &  -1566.6  &  -22.646  &  -1566.597  &  17.7  &  3406.457  &  0.052  &  3410.0  &  -0.104 \\  
 \hline 
J1658-5324  &  334.869  &  -6.625  &  0.88  &  0.2  &  4.9  &  410.0  &  -0.187  &  -1.01  &  -0.183  &  0.83  &  1.938  &  4.283  &  1.94  &  -0.111 \\  
 \hline 
J1905+0400  &  38.095  &  -1.289  &  1.06  &  -3.8  &  -7.3  &  264.2  &  -0.034  &  -1.989  &  -0.029  &  0.647  &  0.797  &  8.118  &  0.798  &  -0.12 \\  
 \hline 
J1543-5149  &  327.921  &  2.479  &  1.15  &  -4.3  &  -4.0  &  486.2  &  -0.382  &  -1.461  &  -0.375  &  1.12  &  1.947  &  5.752  &  1.95  &  -0.15 \\  
 \hline 
J1727-2946  &  357.137  &  2.921  &  1.88  &  0.6  &  0.0  &  36.9  &  -0.034  &  -2.296  &  -0.033  &  0.659  &  1.278  &  5.158  &  1.28  &  -0.181 \\  
 \hline 
J0621+1002  &  200.57  &  -2.013  &  0.42  &  3.23  &  -0.5  &  34.7  &  -0.006  &  -0.415  &  -0.006  &  0.774  &  2.426  &  3.191  &  2.43  &  -0.181 \\  
 \hline 
J1910+1256  &  46.564  &  1.795  &  1.5  &  0.26  &  -7.14  &  200.7  &  -0.039  &  -1.656  &  -0.036  &  1.02  &  0.717  &  14.234  &  0.718  &  -0.197 \\  
 \hline 
J1843-1113  &  22.055  &  -3.397  &  1.26  &  -1.91  &  -3.2  &  541.8  &  -0.281  &  -1.373  &  -0.273  &  1.33  &  1.437  &  9.256  &  1.44  &  -0.214 \\  
 \hline 
J0824+0028  &  223.577  &  20.789  &  1.69  &  -4.3  &  -9.2  &  101.4  &  -0.151  &  -4.199  &  -0.147  &  0.765  &  1.816  &  4.213  &  1.82  &  -0.238 \\  
 \hline 
J1802-2124  &  8.382  &  0.611  &  0.76  &  -0.85  &  4.8  &  79.1  &  -0.045  &  -1.225  &  -0.044  &  0.494  &  2.514  &  1.965  &  2.52  &  -0.244 \\  
 \hline 
J1857+0943  &  42.29  &  3.06  &  1.18  &  -2.665  &  -5.413  &  186.5  &  -0.062  &  -1.095  &  -0.06  &  0.664  &  1.226  &  5.416  &  1.23  &  -0.332 \\  
 \hline 
J1643-1224  &  5.669  &  21.218  &  1.2  &  5.966  &  3.88  &  216.4  &  -0.086  &  -1.646  &  -0.083  &  0.965  &  1.315  &  7.337  &  1.32  &  -0.361 \\  
 \hline 
J2205+6012$^\epsilon$  &  103.686  &  3.696  &  3.53  &  -4.318  &  -3.082  &  414.0  &  -0.339  &  -0.889  &  -0.336  &  1.27  &  0.65  &  19.526  &  0.653  &  -0.394 \\  
 \hline 
J1801-1417  &  14.546  &  4.162  &  1.1  &  -10.89  &  -3.0  &  275.9  &  -0.04  &  -4.419  &  -0.028  &  0.688  &  0.74  &  9.297  &  0.743  &  -0.403 \\  
 \hline 
J1811-2405  &  7.073  &  -2.559  &  1.83  &  0.53  &  0.0  &  375.9  &  -0.195  &  -2.159  &  -0.186  &  1.47  &  0.982  &  14.976  &  0.986  &  -0.447 \\  
 \hline 
J1536-4948  &  328.198  &  4.789  &  0.98  &  -7.3  &  -2.7  &  324.7  &  -0.223  &  -1.837  &  -0.217  &  0.965  &  2.129  &  4.532  &  2.14  &  -0.493 \\  
 \hline 
J1804-2717  &  3.505  &  -2.736  &  0.81  &  2.56  &  -17.0  &  107.0  &  -0.047  &  -6.673  &  -0.04  &  0.587  &  1.917  &  3.062  &  1.93  &  -0.681 \\  
 \hline 
J0117+5914  &  126.283  &  -3.457  &  1.77  &  -3.0  &  18.0  &  9.9  &  -56.852  &  -14.228  &  -56.85  &  17.4  &  109.186  &  1.594  &  110.0  &  -0.74 \\  
 \hline 
J1751-2857  &  0.646  &  -1.124  &  1.09  &  -7.4  &  -4.3  &  255.4  &  -0.073  &  -3.169  &  -0.065  &  1.99  &  1.181  &  16.853  &  1.19  &  -0.775 \\  
 \hline 
J2145-0750  &  47.777  &  -42.084  &  0.83  &  -9.49  &  -9.09  &  62.3  &  -0.012  &  -2.065  &  -0.01  &  0.893  &  1.248  &  7.156  &  1.26  &  -0.962 \\  
 \hline 
J1744-1134  &  14.794  &  9.18  &  0.41  &  18.802  &  -9.415  &  245.4  &  -0.054  &  -4.69  &  -0.042  &  0.977  &  2.583  &  3.783  &  2.61  &  -1.046 \\  
 \hline 
J1045-4509  &  280.851  &  12.254  &  0.59  &  -6.09  &  5.18  &  133.8  &  -0.032  &  -0.319  &  -0.031  &  0.961  &  2.087  &  4.605  &  2.11  &  -1.087 \\  
 \hline 
J2234+0944  &  76.28  &  -40.438  &  0.8  &  6.96  &  -32.17  &  275.7  &  -0.153  &  -19.566  &  -0.099  &  0.858  &  1.909  &  4.495  &  1.93  &  -1.097 \\  
 \hline 
J1708-3506  &  350.47  &  3.124  &  3.32  &  -5.3  &  -2.0  &  222.0  &  -0.056  &  -6.912  &  -0.041  &  0.57  &  0.33  &  17.282  &  0.335  &  -1.547 \\  
 \hline 
J1745-0952  &  16.371  &  9.895  &  0.23  &  -21.2  &  11.0  &  51.6  &  -0.025  &  -3.333  &  -0.023  &  0.472  &  7.388  &  0.639  &  7.52  &  -1.756 \\  
 \hline 
J1810-2005  &  10.545  &  -0.563  &  3.51  &  0.0  &  17.0  &  30.5  &  -0.014  &  -29.49  &  -0.005  &  0.917  &  0.285  &  32.186  &  0.292  &  -2.43 \\  
 \hline 
J0711-6830  &  279.531  &  -23.28  &  0.11  &  -15.565  &  14.18  &  182.1  &  -0.049  &  -1.009  &  -0.048  &  0.696  &  11.854  &  0.587  &  12.3  &  -3.624 \\  
 \hline 
J0437-4715  &  253.394  &  -41.963  &  0.16  &  121.439  &  -71.475  &  173.7  &  -0.173  &  -76.716  &  -0.04  &  0.69  &  7.609  &  0.907  &  7.96  &  -4.413 \\  
 \hline  
J2124-3358  &  10.925  &  -45.438  &  0.44  &  -14.107  &  -50.34  &  202.8  &  -0.085  &  -28.318  &  -0.027  &  0.707  &  2.121  &  3.333  &  2.34  &  -9.346 \\  
 \hline 
J1603-7202  &  316.63  &  -14.496  &  3.4  &  -2.451  &  -7.357  &  67.4  &  -0.007  &  -2.798  &  -0.005  &  0.516  &  0.209  &  24.745  &  0.254  &  -17.902 \\  
 \hline 
J2129-5721  &  338.005  &  -43.57  &  7.0  &  9.294  &  -9.559  &  268.4  &  -0.15  &  -23.238  &  -0.088  &  0.731  &  0.208  &  35.079  &  0.262  &  -20.462 \\  
 \hline 
J0740+6620  &  149.73  &  29.599  &  1.14  &  -10.32  &  -30.87  &  346.5  &  -0.146  &  -29.235  &  -0.045  &  0.626  &  0.807  &  7.757  &  1.06  &  -23.869 \\  
 \hline 
J1946+3417  &  69.294  &  4.713  &  6.94  &  -7.68  &  4.58  &  315.4  &  -0.031  &  -8.239  &  -0.005  &  1.15  &  0.049  &  237.089  &  0.065  &  -25.72 \\  
 \hline 
J2234+0611  &  72.991  &  -43.006  &  1.5  &  25.3  &  9.71  &  279.6  &  -0.094  &  -24.598  &  -0.025  &  0.761  &  0.51  &  14.926  &  0.824  &  -38.125 \\  
 \hline 
J1125-6014$^\epsilon$  &  292.504  &  0.894  &  1.4  &  11.106  &  -13.038  &  380.2  &  -0.054  &  -9.002  &  -0.02  &  0.822  &  0.415  &  19.822  &  0.711  &  -41.674 \\  
 \hline 
J1017-7156  &  291.558  &  -12.553  &  3.5  &  -7.41  &  6.871  &  427.6  &  -0.04  &  -5.436  &  -0.017  &  0.879  &  0.146  &  60.079  &  0.254  &  -42.399 \\  
 \hline  
 \hline   
\end{longtable}
\end{landscape}

\end{appendix}

\end{document}